\documentclass[preprintnumbers,amsmath,amssymb,pra]{revtex4}
\usepackage[latin1]{inputenc}
\usepackage{dcolumn}
\usepackage{bm}
\usepackage{graphicx}

\begin{document}

\title{Unitary transfer of entanglement in multipartite two-level systems}
\author{Riccardo Messina}
\author{Anna Napoli}
\author{Antonino Messina}\affiliation{Dipartimento di Scienze Fisiche ed Astronomiche, Universit\`{a} di Palermo,\\Via Archirafi 36, 90123 Palermo, Italy}

\date{October 28, 2004}

\begin{abstract}
The dynamics of a system composed by two pairs of dipolarly coupled two-level atoms is exactly studied. We show that the initial entanglement stored in a couple of atoms not directly interacting is fully transferred to the other pair in a periodic
way. The observability of this phenomenon in laboratory is briefly discussed both in terms of its temporal scale and of its stability against uncertainties in the geometrical parameters defining the physical system.
\end{abstract}

\maketitle

\section{Introduction}

A multipartite quantum system is by definition a system composed by two or more quantum subsystems. The linear character of Quantum Mechanics implies the existence, in the Hilbert space $\mathcal{E}$ of the total system, of states describing the
occurrence of correlations between a specific pair of subsystems, say $A$ and $B$, even in absence of any mutual interaction coupling them. From a mathematical point of view, such states of $\mathcal{E}$ are those exhibiting the property of being
unfactorizable into the product of two states $|\psi_1\rangle$ and $|\psi_2\rangle$ belonging to any couple of distinct Hilbert spaces $\mathcal{E}_1$ and $\mathcal{E}_2$ respectively, such that $\mathcal{E}=\mathcal{E}_1\otimes\mathcal{E}_2$ and satisfying the further condition that any observable relative to the subsystem $A$ ($B$) acts only upon $\mathcal{E}_1$ ($\mathcal{E}_2$).

It is common to refer to such physical situations saying that the multipartite system is characterized, in the states under scrutiny, by the presence of binary entanglement between $A$ and $B$. The occurrence of entanglement in a multipartite system may of course be associated to physical situations more complex than the one here represented. For example, there exist states where more than one couple of subsystems turn out to be entangled, as well as more complex situations wherein the extension of this peculiar quantum notion to more than two subsystems is currently under debate.

Several experiments aimed at producing entangled states between two atoms \cite{Atomi1,Atomi2,Atomi3} or two photons \cite{Fotoni} have been successfully performed thus providing a positive test of fundamental aspects of Quantum Theory. Over the last years a further interest in generating such a non-classical condition stems from the possibility of considering entanglement as an effective resource in the research area of Quantum Computing \cite{QC} both for storing and for transmitting information.

In this paper we introduce a simple multipartite system composedof four distinguishable identical two-level atoms. Our aim is to provide an example of a simple mechanism for realizing a unitary transfer of entanglement to a couple of non\,-\,interacting atoms. More in detail, we show that, if at $t=0$ the system is prepared in a state where only two atoms among the four are mutually entangled, as a consequence of a strategically chosen interaction mechanism and in absence of any conditional measurement act, the system dynamics drives the complete passage of the initial entanglement towards the pair of atoms disentangled at $t=0$, fully decorrelating at the same time the two initially entangled atoms.

This paper is organized as follows. The next Section describes the physical system and introduces the pertinent hamiltonian model. The dynamics of the system is studied in Section \ref{Dyn}, wherein our main result, that is the possibility of unitarily transferring entanglement, is demonstrated. Some conclusive remarks as well as comments on the experimental feasibility of our proposal are briefly presented in the last Section.

\section{The physical system and its hamiltonian model}

Let us consider four identical two-level atoms named $A_1$, $A_2$, $A_3$ and $A_4$. Let $\hbar\omega$ be the energy separation between the excited ($|+\rangle_i$) and ground ($|-\rangle_i$) states of the $i$-th $(i=1,2,3,4)$ atom. We assume that the only possible interatomic couplings are dipole-dipole interactions between $A_1$ and $A_2$ and between $A_3$ and $A_4$. This could be justified provided that the couples $(A_1,A_2)$ and $(A_3,A_4)$ are located far away from each other. The Hamiltonian of our system is thus
\begin{equation}\label{Hamilt}H\equiv H_0+H_{12}^I+H_{34}^I,\end{equation}
where $H_0$ is the sum of the unperturbated Hamiltonians of the four atoms, whilst $H_{12}^I$ and $H_{34}^I$ are the terms which describe the dipole-dipole interactions. The unperturbated Hamiltonian of $A_i$ can be written in the form
\begin{equation}H_0^i=\frac{\hbar\omega}{2}\sigma_z^{(i)},\end{equation}
$\sigma_z^{(i)}$ being the Pauli operator for the $i$-th two-level atom. The form of the interaction terms is assumed to be \cite{SemLeo,Leo}
\begin{equation}H_{ij}^I=\hbar\eta(\sigma_+^{(i)}\sigma_-^{(j)}+\mathrm{h.c.}),\end{equation}
where $(ij)=(12)$ or $(ij)=(34)$ and $\sigma_\pm^{(i)}=|\mp\rangle_i\,{}_i\langle\pm|$. The coupling strength $\eta$ may be expressed in terms of the geometric distance $R$ between atoms $(A_1,A_2)$ or $(A_3,A_4)$ in the following form:
\begin{equation}\label{Costante}\eta=\frac{3}{4}\frac{\Gamma_0c^3}{w^3R^3}(1-\cos^2\alpha),\end{equation}
where $\alpha$ is the angle between $(\mathbf{r}_1-\mathbf{r}_2)=(\mathbf{r}_3-\mathbf{r}_4)$ and the atomic transition dipole moment $\mathbf{d}$, $\mathbf{r}_i$ being the $i$-th atomic position vector. Finally, $\Gamma_0$ is the spontaneous emission rate of each atom in free space.

It is convenient to cast our hamiltonian model defined in eq.\eqref{Hamilt} in the form
\begin{equation}H\equiv H_{12}+H_{34}\end{equation}
with
\begin{equation}H_{ij}=\frac{\hbar\omega}{2}(\sigma_z^
{(i)}+\sigma_z^{(j)})+\hbar\eta(\sigma_+^{(i)}\sigma_-^{(j)}+\sigma_i^{(i)}\sigma_+^{(j)})\qquad
(i,j)=(1,2)\quad\mathrm{or}\quad(3,4)\end{equation}
because in this way it appears evidently symmetric against the exchange of the couple of indexes $(1,2)$ with the couple $(3,4)$.

Taking into account that $H_{12}$ commutes with $(\sigma_z^{(1)}+\sigma_z^{(2)})$ and with $(\overrightarrow{\sigma}_{1}+\overrightarrow{\sigma}_{2})^2$, it is possible to convince oneself that a unitary operator accomplishing the transformation of $H_{12}$ into the diagonal Hamiltonian $\widetilde{H}_{12}\equiv U_{12}H_{12}U_{12}^\dag$ may be given by
\begin{equation}U_{12}\equiv\exp\left\{\frac{\pi}{4}\Bigl(\sigma_+^{(1)}\sigma_-^{(2)}-\sigma_-^{(1)}\sigma_+^{(2)}\Bigr)\right\}.\end{equation}
This operator transforms indeed only the interaction term, so that
\begin{equation}\widetilde{H}_{12}=\frac{\hbar\omega}{2}(\sigma_z^
{(1)}+\sigma_z^{(2)})+\hbar\eta
U_{12}(\sigma_+^{(1)}\sigma_-^{(2)}+\sigma_-^{(1)}\sigma_+^{(2)})U_{12}^\dag.\end{equation}
To transform the operators $\sigma_+^{(1)}\sigma_-^{(2)}$ and $\sigma_-^{(1)}\sigma_+^{(2)}$ we make use of the well-known Baker-Campbell-Hausdorff formula obtaining
\begin{equation}\label{Trasf1}U_{12}\sigma_+^{(1)}\sigma_-^{(2)}U_{12}^\dag=\sigma_+^{(1)}\sigma_-^{(2)}+\frac{1}{4}(\sigma_z^
{(1)}-\sigma_z^{(2)})-\frac{1}{2}(\sigma_+^{(1)}\sigma_-^{(2)}+\sigma_-^{(1)}\sigma_+^{(2)})\end{equation}
and
\begin{equation}\label{Trasf2}U_{12}\sigma_-^{(1)}\sigma_+^{(2)}U_{12}^\dag=\sigma_-^{(1)}\sigma_+^{(2)}+\frac{1}{4}(\sigma_z^
{(1)}-\sigma_z^{(2)})-\frac{1}{2}(\sigma_+^{(1)}\sigma_-^{(2)}+\sigma_-^{(1)}\sigma_+^{(2)}).\end{equation}

Exploiting the additive form of $H$ with respect to the two subsystems $(A_1,A_2)$ and $(A_3,A_4)$ as well as its previously mentioned exchange symmetry, if we are able to find a unitary operator $U_{12}$ accomplishing the diagonalization process of $H_{12}$, then we can diagonalize the entire Hamiltonian $H$ using the operator $U\equiv U_{12}U_{34}$, where $U_{34}$ may be simply derived from $U_{12}$ putting into its expression $(3,4)$ in place of $(1,2)$. Extending eqs.\eqref{Trasf1} and \eqref{Trasf2} to the subsystem $(A_3,A_4)$ with the help of the unitary operator $U_{34}$, the transformed Hamiltonian
\begin{equation}\widetilde{H}\equiv UHU^\dag=U_{12}U_{34}HU_{34}^\dag U_{12}^\dag\end{equation}
assumes the form
\begin{equation}\widetilde{H}=\widetilde{H}_{12}+\widetilde{H}_{34}\end{equation}
where
\begin{equation}\widetilde{H}_{ij}\equiv\frac{\hbar\omega}{2}(\sigma_z^
{(i)}+\sigma_z^{(j)})+\frac{\hbar\eta}{2}(\sigma_z^
{(i)}-\sigma_z^{(j)})\qquad
(i,j)=(1,2)\quad\mathrm{or}\quad(3,4)\end{equation}
The new Hamiltonian $\widetilde{H}$ is, as requested, manifestly diagonal.

\section{Unitary transfer of entanglement}\label{Dyn}

Let us suppose that at the time instant $t=0$ our system has been prepared in the initial state
\begin{equation}\label{Stiniz}\begin{split}|\psi(0)\rangle&=|-\rangle_1\Bigl(\cos\theta|+\rangle_2|-\rangle_3+e^{i\varphi}\sin\theta|-\rangle_2|+\rangle_3\Bigr)|-\rangle_4=\\
&=\cos\theta|-\rangle_1|+\rangle_2|-\rangle_3|-\rangle_4+e^{i\varphi}\sin\theta|-\rangle_1|-\rangle_2|+\rangle_3|-\rangle_4\equiv\\
&\equiv\cos\theta|-+--\rangle+e^{i\varphi}\sin\theta|--+-\rangle,\\\end{split}\end{equation}
where $\theta$ and $\varphi$ are real parameters. For any $\theta\neq k\frac{\pi}{2}$ $(k\in\mathbb{Z})$, the state \eqref{Stiniz} describes the occurrence of entanglement only between $A_2$ and $A_3$, meaning that each of the two atoms $A_1$
and $A_4$ turns out to be disentangled form the other three atoms of the system.

The time evolution of this initial state of the system may be obtained exploiting the knowledge of the unitary operator $U$ diagonalizing $H$, since
\begin{equation}\label{Evol}|\psi(t)\rangle=e^{-i\frac{H}{\hbar}t}|\psi(0)\rangle=U^\dag Ue^{-i\frac{H}{\hbar}t}U^\dag U|\psi(0)\rangle=U^\dag
e^{-i\frac{\widetilde{H}}{\hbar}t}U|\psi(0)\rangle.\end{equation}
To evaluate the action of $U$ upon $|\psi(0)\rangle$ we proceed observing that, by definition,
\begin{equation}U_{12}\equiv\exp\left\{\frac{\pi}{4}\Bigl(\sigma_+^{(1)}\sigma_-^{(2)}-\sigma_-^{(1)}\sigma_+^{(2)}\Bigr)\right\}\equiv\sum_{n=0}^{+\infty}\frac{\Bigl[\frac{\pi}{4}\Bigl(\sigma_+^{(1)}\sigma_-^{(2)}-\sigma_-^{(1)}\sigma_+^{(2)}\Bigr)\Bigr]^n}{n!}.\end{equation}
By simple calculations it is not difficult to find the following polynomial expression of $U_{12}$:
\begin{equation}\label{U12}\begin{split}&U_{12}=I+\frac{1}{\sqrt{2}}\,X+\Bigl(1-\frac{1}{\sqrt{2}}\Bigr)X^2=\\
&=I+\frac{1}{\sqrt{2}}\Bigl(\sigma_+^{(1)}\sigma_-^{(2)}-\sigma_-^{(1)}\sigma_+^{(2)}\Bigr)-\left(1-\frac{1}{\sqrt{2}}\right)\Bigl(\sigma_+^{(1)}\sigma_-^{(1)}\sigma_-^{(2)}\sigma_+^{(2)}+\sigma_-^{(1)}\sigma_+^{(1)}\sigma_+^{(2)}\sigma_-^{(2)}\Bigr).\\\end{split}\end{equation}

In view of the simple connection between $U_{12}$ and $U_{34}$, eq.\eqref{U12} helps us to write down immediately a polynomial expression of $U=U_{12}U_{34}$, by which we succeed in evaluating the transformed initial condition $U|\psi(0)\rangle$ appearing in eq.\eqref{Evol} as follows:
\begin{equation}\begin{split}U|\psi(0)\rangle &=\frac{1}{\sqrt{2}}\Bigl[\cos\theta\Bigl(|+---\rangle+|-+--\rangle\Bigr)+\\
&\quad+e^{i\varphi}\sin\theta\Bigl(|--+-\rangle-|---+\rangle\Bigr)\Bigr].\\\end{split}\end{equation}
We are now ready to follow up the steps suggested by eq.\eqref{Evol}, getting the following explicit expression for $|\psi(t)\rangle$:
\begin{equation}\label{Stevoluto}\begin{split}|\psi(t)\rangle&=e^{i\omega t}\Bigl\{\cos\theta\Bigl[-i\sin(\eta t)|+---\rangle+\cos(\eta t)|-+--\rangle\Bigl]+\\
&\quad+e^{i\varphi}\sin\theta\Bigl[-i\sin(\eta
t)|---+\rangle+\cos(\eta
t)|--+-\rangle\Bigr]\Bigr\}.\\\end{split}\end{equation}

It is of relevance to observe that, as a consequence of the initial entanglement between $A_2$ and $A_3$ as well as of the assumed interaction mechanism within the subsystems $(A_1,A_2)$ and $(A_3,A_4)$, there exist intervals of time during which the system is able to exhibit an entanglement condition involving all the four atoms. This means that, apart from isolated time instants, the state $|\psi(t)\rangle$ cannot be factorized into the product of two states pertaining to two subsystems
partitioning the global one. Had we renounced to inject at $t=0$ an entanglement between an atom of the subsystem $(A_1,A_2)$ and an atom of the other one, the consequent evolution under the action of our hamiltonian model would have assured the presence of entanglement only in the subsystem $(A_1,A_2)$, $(A_3,A_4)$, or both at most. As a consequence, our choice of initially entangling one atom in $(A_1,A_2)$ with one in $(A_3,A_4)$ only (in our case $A_2$ and $A_3)$, appears to be the simplest necessary starting point to legitimate the searching of time instants at which the only couple of entangled atoms is $(A_1,A_4)$. This amounts at looking for time instants at which the evolution drives the four-atom system into a state structurally analogous to the initial one, meaning that a couple of atoms (in our case $A_1$ and $A_4$) is in an entangled state whilst the remaining two atoms are in a factorized one. It is not difficult to persuade oneself that the constants of motion imposed by our hamiltonian model
restrict the class of states describing the desired unitary structural transfer of entanglement to the set of states having the form
\begin{equation}\label{Cercato}|\psi_S\rangle=\cos\lambda|+---\rangle+e^{i\zeta}\sin\lambda|---+\rangle,\end{equation}
where $\lambda$ and $\zeta$ are real parameters.

Thus our problem consists in establishing whether time instants exist at which $|\psi(t)\rangle$ given by eq.\eqref{Stevoluto} does belong to the set expressed by eq.\eqref{Cercato} in correspondence to appropriate values of the parameters $\lambda$ and $\zeta$. To solve this question we pose the following equation in $t$
\begin{equation}A(t)\equiv|\langle\psi_S|\psi(t)\rangle|^2=1\end{equation}
that is explicitly
\begin{equation}\label{Eqdef}\sin^2(\eta t)\Bigl[\cos^2\theta\cos^2\lambda+\sin^2\theta\sin^2\lambda+\frac{1}{2}\sin(2\theta)\sin(2\lambda)\cos(\varphi-\zeta)\Bigr]=1.\end{equation}
Taking into account that
\begin{equation}0\leq A(t)\leq\Bigl(|\cos\theta\cos\lambda|+|\sin\theta\sin\lambda|\Bigr)^2\end{equation}
and that, in addition,
$(|\cos\theta\cos\lambda|+|\sin\theta\sin\lambda|)^2$ coincides
with $\cos^2(\theta-\lambda)$ or $\cos^2(\theta+\lambda)$, then
eq.\eqref{Eqdef} necessarily requires
\begin{equation}\label{Cond1}\lambda=\pm\theta\end{equation}
apart from additive multiples of $\pi$, leading to physically undistinguishable states having the form of $|\psi_S\rangle$.

Inserting condition \eqref{Cond1} into eq.\eqref{Eqdef} we easily get the following further necessary condition on the parameter $\zeta$:
\begin{equation}\label{Cond2}
\begin{cases}
\zeta=\varphi & \text{if}\quad\lambda=\theta\\
\zeta=\varphi+\pi & \text{if}\quad\lambda=-\theta\\
\end{cases}
\end{equation}
Thus, in conclusion, exploiting eqs.\eqref{Cond1} and \eqref{Cond2}, eq.\eqref{Eqdef} is simply equivalent to the equation $\sin^2(\eta t)=1$ implying
\begin{equation}t_n=\frac{\pi}{2\eta}(2n+1)\qquad n\in\mathbb{N}.\end{equation}
Summing up, we have proved that, at all such time instants $t_n$, the initial state evolves into the state
\begin{equation}\label{Sttrasf}|\psi(t_n)\rangle\equiv|\psi_T\rangle=\cos\theta|+---\rangle+e^{i\varphi}\sin\theta|---+\rangle,\end{equation}
sharing its structure, under the entanglement point of view, with $|\psi(0)\rangle$.

From an experimental point of view, the first time instant $t_0=\frac{\pi}{2\eta}$ plays a special role in view of the fact that it gives the temporal scale for the generation under scrutiny, to be compared with the appearance of countering effects
traceable back to unavoidable coupling with environment.

The periodicity related to the temporal behavior of the system at the time instants $t_n$ suggests to look for a possible periodic appearance of the initial condition in the dynamical evolution of the system. We find indeed that at the intermediate time instants between the $t_n$, that means at the time instants
\begin{equation}\label{Tinter}t_m=\frac{\pi}{\eta}(m+1)\qquad m\in\mathbb{N}.\end{equation}
the system effectively passes through a state coincident with $|\psi(0)\rangle$ apart from a global constant phase factor. Our main result is therefore that the time evolution of our four-atom system has the form of an oscillation between the initial state $|\psi(0)\rangle$, incorporating the entanglement of the couple of atoms $(A_2,A_3)$, and the state $|\psi_T\rangle$ given by eq.\eqref{Sttrasf}, clearly describing that entanglement has been transferred to the couple $(A_1,A_4)$. The expressions for the first time instant at which the state of the system is $|\psi_T\rangle$, and of the period of the oscillation have been found to be
\begin{equation}t_0=\frac{\pi}{2\eta}\qquad T=\frac{\pi}{\eta}.\end{equation}

\section{Discussion and conclusive remarks}

The results reported in the previous Section rely on some assumptions concerning both the geometry of the model and the physical mechanisms of energy exchange within the four-atom system. While dipole-dipole interactions are very often used to modelize microscopic couplings between two two-level systems \cite{Modellidipolo}, in order to fully appreciate the reliability of our conclusions it is worth analyzing their stability against possibile geometric uncertainties. We have indeed assumed that the relative position $(\mathbf{r}_1-\mathbf{r}_2)$ between $A_1$ and $A_2$ is exactly coincident with the one, $(\mathbf{r}_3-\mathbf{r}_4)$, between $A_3$ and $A_4$, leading in this way to put equal coupling strengths $\eta$ within the two couples of atoms. Indicating then the coupling strength $\eta_{12}$ between $A_1$ and $A_2$ with $\eta$ and, more realistically, putting $\eta_{34}$ equal to $\eta+\delta\eta$, we may start again the study of the time evolution of the system, getting the time evolved state
\begin{equation}\begin{split}|\widetilde{\psi}(t)\rangle&=e^{i\omega t}\Bigl\{\cos\theta\Bigl[-i\sin(\eta t)|+---\rangle+\cos(\eta t)|-+--\rangle\Bigl]+\\
&\quad+e^{i\varphi}\sin\theta\Bigl[-i\sin\Bigl[(\eta+\delta\eta)t\Bigr]|---+\rangle+\cos\Bigl[(\eta+\delta\eta)t\Bigr]|--+-\rangle\Bigr]\Bigr\}.\\\end{split}\end{equation}
The analysis of at what extent our results may be claimed stable against small variations affecting the geometrical parameters of the system, may be performed calculating, at $t=t_0=\frac{\pi}{2\eta}$, the fidelity of $|\widetilde{\psi}(t_0)\rangle$ with respect to the state $|\psi_T\rangle$ given by eq.\eqref{Sttrasf}, obtaining
\begin{equation}\begin{split}|\langle\psi_T|\widetilde{\psi}(t_0)\rangle|&=\cos^2\theta+\sin^2\theta\sin(\eta_{34}t)=\cos^2\theta+\sin^2\theta\cos\Bigl(\frac{\pi}{2}\frac{\delta\eta}{\eta}\Bigr)\simeq\\
&\simeq
1-\frac{1}{2}\Bigl(\frac{\delta\eta}{\eta}\Bigr)^2\sin^2\theta\geq
1-\frac{1}{2}\Bigl(\frac{\delta\eta}{\eta}\Bigr)^2.\\\end{split}\end{equation}

Although the main result of this paper is the possibility of transferring the storage of a given amount of entanglement from a subsystem to another one, we wish however to underline that it is in the current experimental reach both to build up prefixed geometrical configurations of a set of atoms, as well as to prepare states of a selected couple of two-level atoms like the initial one we have chosen for the couple $(A_2,A_3)$, as
expressed by eq.\eqref{Stiniz} \cite{Atomi1,Atomi2,Atomi3}.

We wish to conclude pointing out that, to observe in laboratory the periodic passage of entanglement in accordance with the results found in this paper, it is necessary to protect the coherent dynamics of the system against any source of noise
effects. In other words, the realization of our physical scenario has to be conceived so to guarantee temporal scales of the coherent phenomenon much shorter than the characteristic times of environmental effects. In the framework of cavity quantum
electrodynamics, it is possible to reach values of $\eta$ of the order of $10^6Hz$ \cite{Modellidipolo,Pellizzari} determining the appearance of the first transfer of entanglement after $t_0\sim10^{-6}s$ to be compared with the decoherence temporal scale estimable of the order of $10^{-5}s$.

\end{document}